\documentclass[twocolumn]{jpsj3}

\usepackage{color}
\usepackage{times}
\usepackage{graphicx}

\title{Three-Channel Kondo Effect Emerging from Ho Ions}

\author{Takashi Hotta}

\inst{Department of Physics, Tokyo Metropolitan University,
Hachioji, Tokyo 192-0397, Japan}

\recdate{\today}

\abst{
We illustrate the emergence of the three-channel Kondo effect
in a seven-orbital impurity Anderson model hybridized with
$\Gamma_7$ and $\Gamma_8$ conduction electron bands
employing a numerical renormalization group method.
In particular, we focus on the case with ten local $f$ electrons
corresponding to a Ho$^{3+}$ ion.
By investigating the dependence of the low-temperature behavior of
the $f$-electron entropy on the crystalline electric field potential and
hybridization, we found a residual entropy of $\log \phi$
with the golden ratio $\phi=(1+\sqrt{5})/2$,
characteristic of the three-channel Kondo effect,
for the wide range of parameters of the local $\Gamma_5$ triplet
ground state.
To improve our insight of local impurity spin,
we explored a quantum critical point
between three-channel Kondo and Fermi-liquid phases.
Finally, we briefly discussed the cubic Ho compound as a candidate
to observe experimentally the three-channel Kondo effect.
}

\begin{document}
\maketitle

The multi-channel Kondo effect has been discussed for a long time
as a potential source of exotic quantum ground states, such as
the non-Fermi liquid phase, after the pioneering proposal of
the multi-channel Kondo effect concept
by Nozi\`eres and Blandin.\cite{Nozieres}
In particular, the two-channel Kondo effect
has been vigorously researched both in experimental
and theoretical fields. This research intensified after Cox
pointed out that there exist
two screening channels concerning the quadrupole degrees of freedom
of a cubic uranium compound
with non-Kramers doublet ground state.\cite{Cox1,Cox2}
Experimental studies to observe
the two-channel Kondo effect have had significant advances in
observing the signals of the two-channel Kondo effect
in cubic Pr compounds with non-Kramers doublet ground state.
\cite{Sakai,Onimaru1,Onimaru2,Higashinaka,Review}

For the past three decades,
there has been significant attention on the quadrupole Kondo
phenomenon to realize the two-channel Kondo effect,
but it is also important to pay renewed attention to
the magnetic two-channel Kondo effect based on
the original concept by Nozi\`eres and Blandin.
From this revival viewpoint,
the present author has confirmed the emergence of
the two-channel Kondo effect in Nd ions for the wide range of
parameters of the local $\Gamma_6$ ground state.\cite{Hotta1}
This is considered to be the magnetic two-channel Kondo effect.

Here, we have a naive question on the realization of
the multi-channel Kondo phenomena
beyond the two-channel Kondo effect.
For instance, it is intriguing to clarify
the appearance of the three-channel Kondo effect.
Concerning this point, in the three-orbital Anderson model of a single
C$_{60}$ molecule, Leo and Fabrizio discussed
the phase diagram including the three-channel Kondo state.\cite{Fabrizio}
However, research on the three-channel Kondo effect
in $f$-electron systems, including rare-earth and actinide ions,
has not been performed.
The three-channel Kondo effect was not
observed in our previous research on the three-band Anderson model
with light rare-earth ions.
Thus, it is necessary to focus on heavy rare-earth ions,
which have not been regarded as a stage for the multi-channel Kondo effect.

In this study, we propose that the three-channel Kondo effect
can be observed in the cubic Ho compound.
By using a numerical renormalization group (NRG) method,
we analyze a seven-orbital impurity Anderson model hybridized
with $\Gamma_8$ and $\Gamma_7$ conduction electrons
for a case with ten local $f$ electrons corresponding to a Ho$^{3+}$ ion.
Then, we find a residual entropy of $\log \phi$ with the golden ratio
$\phi=(1+\sqrt{5})/2$ as a signal of the three-channel Kondo effect
for the local $\Gamma_5$ triplet ground state.
We also discuss local impurity spin
by investigating quantum critical behavior between
the three-channel Kondo and Fermi-liquid phases.
Finally, we provide a short comment on the candidate material
to detect the three-channel Kondo effect.

For the description of the local $f$-electron model,
first, we define the one $f$-electron state from the eigenstate of
the spin-orbit and crystalline electric field (CEF) potential terms.
Under the cubic CEF potentials,
we find $\Gamma_7$ doublet and $\Gamma_8$ quartet
from $j=5/2$ sextet,
whereas we obtain a $\Gamma_6$ doublet, $\Gamma_7$ doublet,
and $\Gamma_8$ quartet from the $j=7/2$ octet.
With the use of these one-electron states as bases,
the local $f$-electron Hamiltonian is expressed as
\begin{equation}
\label{Hloc}
\begin{split}
  H_{\rm loc} &=\sum_{j, \mu, \tau} (\lambda_j  + B_{j,\mu})
  f_{j \mu \tau}^{\dag} f_{j \mu \tau} +E_{f}n \\
  &+\sum_{j_1\sim  j_4} \sum_{\mu_1 \sim \mu_4}
  \sum_{\tau_1 \sim \tau_4} 
  I^{j_1 j_2, j_3 j_4}_{\mu_1 \tau_1 \mu_2 \tau_2, \mu_3 \tau_3 \mu_4 \tau_4} \\
 &\times  f_{j_1 \mu_1 \tau_1}^{\dag} f_{j_2 \mu_2 \tau_2}^{\dag}
  f_{j_3 \mu_3 \tau_3}  f_{j_4 \mu_4 \tau_4},
\end{split}
\end{equation}
where $f_{j \mu\tau}$ is the annihilation operator of
a localized $f$ electron in the bases of $(j, \mu, \tau)$,
$j$ is the total angular momentum,
$j=5/2$ and $7/2$ are denoted by ``$a$'' and ``$b$'', respectively,
$\mu$ distinguishes the cubic irreducible representation,
$\Gamma_8$ states are distinguished by $\mu=\alpha$ and $\beta$,
while the $\Gamma_7$ and $\Gamma_6$ states are labeled
by $\mu=\gamma$ and $\delta$, respectively,
$\tau$ denotes the pseudo-spin, which distinguishes the degeneracy
concerning the time-reversal symmetry,
$n$ is the local $f$-electron number at an impurity site,
and $E_f$ is the $f$-electron level to control $n$.
In this paper, we set $\hbar=k_{\rm B}=1$ and the energy unit as eV.

Concerning the spin-orbit term, we obtain
$\lambda_a=-2\lambda$ and $\lambda_b=3\lambda/2$,
where $\lambda$ is the spin-orbit coupling of the $f$ electron.
In this study, we set $\lambda=0.265$ for the Ho ions.\cite{spin-orbit}
For the CEF potential term of $j=5/2$,
we obtain $B_{a,\alpha}=B_{a,\beta}=1320 B_4^0/7$
and $B_{a,\gamma} =-2640 B_4^0/7$,
where $B_4^0$ denotes the fourth-order CEF parameter
in the table of Hutchings for the angular momentum $\ell=3$.
\cite{Hutchings}
Note that the sixth-order CEF potential term $B_6^0$
does not appear for $j=5/2$ since the maximum size
of the change of the total angular momentum is less than six.
On the other hand, for $j=7/2$, we obtain
$B_{b,\alpha}=B_{b,\beta}=360B_4^0/7+2880B_6^0$,
$B_{b,\gamma}=-3240B_4^0/7-2160 B_6^0$,
and $B_{b,\delta}=360 B_4^0-3600 B_6^0/7$.
Note that the $B_6^0$ terms appear in this case.
For the Coulomb interaction terms,
we do not show the explicit forms of $I$,
but they are expressed by the four Slater-Condon parameters,
$F^0$, $F^2$, $F^4$, and $F^6$.\cite{Slater}
These values should be determined from experimental results,
but here we simply set the ratio as
$F^0/10=F^2/5=F^4/3=F^6=U$,\cite{Hotta-Harima}
where $U$ indicates the Hund's rule interaction among the $f$ orbitals
and the magnitude is set as unity in this study.

\begin{figure}[t]
\centering
\includegraphics[width=8truecm]{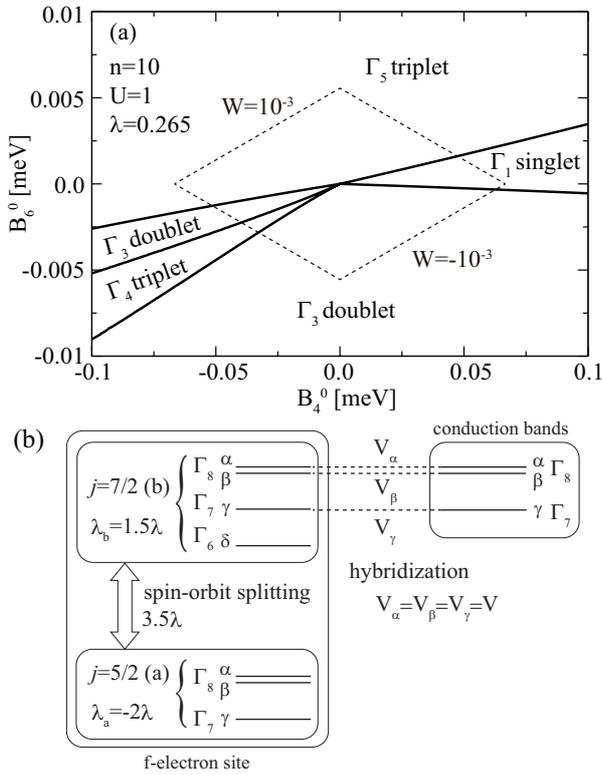}
\caption{(a) Local CEF ground-state phase diagram
on the $(B_4^0, B_6^0)$ plane for $n=10$
with $U=1$ and $\lambda=0.265$.
The dashed rhombus denotes the trajectory of $B_4^0=Wx/15$ and
$B_6^0=W(1-|x|)/180$ for $-1 \le x \le 1$ with $W=\pm 10^{-3}$.
(b) Schematic view of the seven-orbital impurity Anderson model.
The left part indicates the local $f$-electron states
in a $j$-$j$ coupling scheme,
whereas the upper right part denotes three conduction bands
hybridized with the $j=7/2$ states of
the same irreducible representations.
}
\end{figure}

First let us consider the local CEF ground-state phase diagram
for $n=10$.
The ground-state multiplet for $B_4^0=B_6^0=0$ is
characterized by the total angular momentum $J=8$.
Under the cubic CEF potentials, the sept-dectet of $J=8$
is split into four groups as one $\Gamma_1$ singlet,
two $\Gamma_3$ doublets, two $\Gamma_4$ triplets,
and two $\Gamma_5$ triplets.\cite{LLW}
Then, we obtain four kinds of local ground states for $n=10$,
as shown in Fig.~1(a).
Roughly speaking, the $\Gamma_5$ triplet appears widely for $B_6^0>0$,
whereas the $\Gamma_3$ doublet is found for $B_6^0<0$.
In the region of $B_6^0 \approx 0$ and $B_4^0>0$,
the $\Gamma_1$ singlet was stabilized.
For $B_4^0<0$, we find the $\Gamma_3$ doublet and
$\Gamma_4$ triplet in a narrow region
between the $\Gamma_5$ triplet and $\Gamma_3$ doublet.

In the following calculations, we use the parametrization as
$B_4^0=Wx/F(4)$ and $B_6^0=W(1-|x|)/F(6)$,\cite{LLW}
where $x$ specifies the CEF scheme for the $O_{\rm h}$ point group,
while $W$ determines the energy scale of the CEF potentials.
We choose $F(4)=15$ and $F(6)=180$ for $\ell=3$.\cite{Hutchings}
The trajectory of $B_4^0$ and $B_6^0$ for $-1 \le x \le 1$
with a fixed value of $|W|$ ($W>0$ and $W<0$) forms
a rhombus in the planes of $B_4^0$ and $B_6^0$.
The rhombus for $|W|=10^{-3}$ is shown in Fig.~1(a).

To construct the impurity Anderson model,
we further consider the $\Gamma_7$ and $\Gamma_8$ conduction
electron bands hybridized with localized $f$ electrons.
Since we focus on the case of $n=10$,
the $j=5/2$ sextet is considered to be fully occupied and
the Fermi level is situated among the $j=7/2$ octet.
Thus, we consider only the hybridization between the conduction and
$j=7/2$ electrons.
As schematically shown in Fig.~1(b),
the seven-orbital Anderson model is given by
\begin{equation}
H \!=\! \sum_{\mib{k},\mu,\tau} \varepsilon_{\mib{k}}
c_{\mib{k}\mu\tau}^{\dag} c_{\mib{k}\mu\tau}
+ \! \sum_{\mib{k},\mu,\tau} V_{\mu}
(c_{\mib{k}\mu\tau}^{\dag}f_{b\mu\tau}+{\rm h.c.})
+ \! H_{\rm loc},
\end{equation}
where $\varepsilon_{\mib{k}}$ is the dispersion of
the conduction electron with the wave vector $\mib{k}$,
$c_{\mib{k}\mu\tau}$ is the annihilation operator of the
conduction electrons,
and $V_{\mu}$ denotes the hybridization between
the localized and conduction electrons of the $\mu$ orbital.
In this study, we assume
$V_{\alpha}=V_{\beta}=V_{\gamma}=V$.
Note that $V_{\alpha}=V_{\beta}$ from the cubic symmetry,
whereas $V_{\gamma}$ can take a different value from
that of the $\Gamma_8$ conduction band.
The effect of the difference in $V_{\alpha}$ and $V_{\gamma}$
will be discussed in a future study.

In this study, we analyzed the model by employing the NRG method.
\cite{NRG1,NRG2}
We introduced a cut-off $\Lambda$ for the logarithmic discretization
of the conduction band.
Due to limited computer resources,
we kept $M$ low-energy states.
Here, we used $\Lambda=8$ and $M=5,000$.
In the NRG calculation, the temperature $T$ was defined as
$T=D \Lambda^{-(N-1)/2}$,
where $D$ is half the conduction band width,
which was set as 1 eV,
and $N$ is the number of renormalization steps.

\begin{figure}[t]
\centering
\includegraphics[width=8.0truecm]{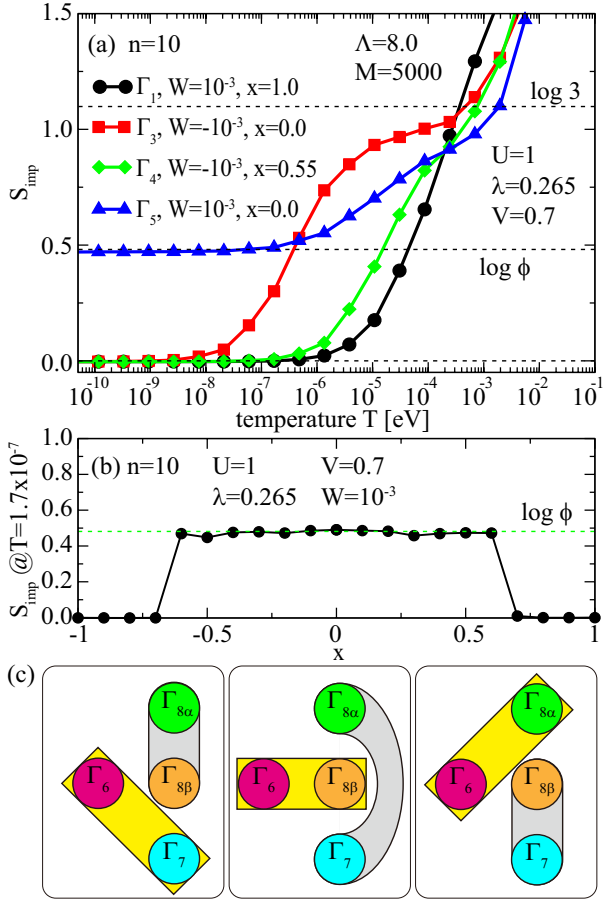}
\caption{(Color online)
(a) Entropies vs. temperature for several sets of $W$ and $x$.
(b) Residual entropies at $T=1.7 \times 10^{-7}$ vs. $x$
for $W=10^{-3}$.
(c) Schematic view of the main components of
the local $\Gamma_5$ triplet state.
The rectangle and oval denote the triplet and singlet pairs,
respectively.
}
\end{figure}

Now, we show the NRG results of $f$-electron entropy.
In Fig.~2(a), we select several results for
$(W,x)=(10^{-3}, 1.0)$ ($\Gamma_1$ singlet),
$(-10^{-3}, 0.0)$ ($\Gamma_3$ doublet),
$(-10^{-3}, 0.55)$ ($\Gamma_4$ triplet),
and $(10^{-3}, 0.0)$ ($\Gamma_5$ triplet) with $V=0.7$.
For the $\Gamma_1$ singlet, we obtain the CEF singlet state.
In fact, the residual entropy becomes zero monotonically.
For the $\Gamma_4$ triplet, we find a tiny shoulder structure
with an entropy smaller than $\log 3$ at around $T \sim 10^{-4}$,
suggesting a remnant of a local triplet state,
but at low temperatures, we obtain the Fermi-liquid phase.
For the $\Gamma_3$ doublet, we observe the shoulder structure
for an entropy smaller than $\log 3$.
The ground state is the $\Gamma_3$ doublet,
but there exists an excited state of a $\Gamma_4$ triplet
with a very small excitation energy.\cite{Hotta2005,Hotta2007}
Due to the hybridization effect, the role of the $\Gamma_4$ triplet
becomes significant.
Subsequently, a signal of a local triplet state appears.
Note that at low temperatures, we also arrive at the Fermi-liquid phase.

Let us discuss the case of the $\Gamma_5$ triplet
where a residual entropy of $\log \phi$ is observed
at low temperatures
with the golden ratio $\phi=(1+\sqrt{5})/2$.
The analytic value of the residual entropy $S_{\rm ana}$
for the multi-channel Kondo effect is given by~\cite{Affleck}
\begin{equation}
\label{eq:Sana}
S_{\rm ana}=\log \frac{\sin [(2S +1)\pi/(n_{\rm c}+2)]}
{\sin [\pi/(n_{\rm c}+2)]},
\end{equation}
where $S$ denotes the local impurity spin and
$n_{\rm c}$ indicates the number of channels.
In the present case with $n_{\rm c}=3$,
we obtain $S_{\rm ana}=\log \phi$
for both the cases of $S=1/2$ and $1$.
Thus, in any case, the residual entropy $\log \phi$ is
considered to be characteristic of the three-channel Kondo effect.

To clarify the range of the parameter region for
the appearance of the three-channel Kondo phenomenon,
we repeat the NRG calculations by changing the values of
$x$ from $-1.0$ to $1.0$ with step of $0.1$
while the other parameters are fixed.
In Fig.~2(b), we depict the residual entropy
at $T=1.7 \times 10^{-7}$ as a function of $x$.
We observe an entropy plateau of $\log \phi$
for a wide range of parameters in the region of the
$\Gamma_5$ triplet, as shown in Fig.~1(a).
Note that the $\Gamma_5$ triplet state also appears
in the region of $-1.0 \le x \le -0.7$.
However, owing to the effect of the first excited state of
the $\Gamma_3$ doublet with a small excitation energy,
the three-channel Kondo phase is considered to be suppressed.
In conclusion, the three-channel Kondo effect appears
in the region of the local $\Gamma_5$ triplet state.

For $n_{\rm c}=3$, the value of $S$ is not determined
only from the residual entropy,
although $S=1$ is expected in the $\Gamma_5$ triplet state.
To clarify the origin of the local impurity spin,
we briefly discuss the local ground state.
Since the $j=5/2$ states are fully occupied
in the main components of the local ground state for $n=10$,
we accommodate four electrons in the $j=7/2$ states.
From the numerical diagonalization of $H_{\rm loc}$,
we find that the main components of the local ground states
are expressed by the combination of the pseudo-spin triplet
and singlet pairs, as schematically shown in Fig.~2(c).
Note that the ground states include other components
obtained by exchanging the triplet and singlet pairs.
The existence of the pseudo-spin triplet pair in the
local ground states seems to be consistent with
the magnetic origin of $S=1$.

To confirm the value of $S$, we explore a quantum critical point (QCP)
between the three-channel Kondo and Fermi-liquid phases.
It has been recognized that the QCP appears at the transition
between the screened Kondo and local singlet phases,
characterized by the residual entropy of $0.5 \log 2$.
\cite{Koga1,Koga2,Kusunose1,Kusunose2,OSakai,Koga3,Koga4,Miyake1,
Fabrizio1,Fabrizio2,Miyake2,Koga5,Miyake3,Miyake4,Sela,
Shiina1,Shiina2,Hotta2,Koga6,Hotta3}
Quite recently, the present author clarified that the QCP
between the two-channel Kondo and Fermi-liquid phases
is characterized by $\log \phi$.\cite{Hotta2020}

\begin{figure}[t]
\centering
\includegraphics[width=8.0truecm]{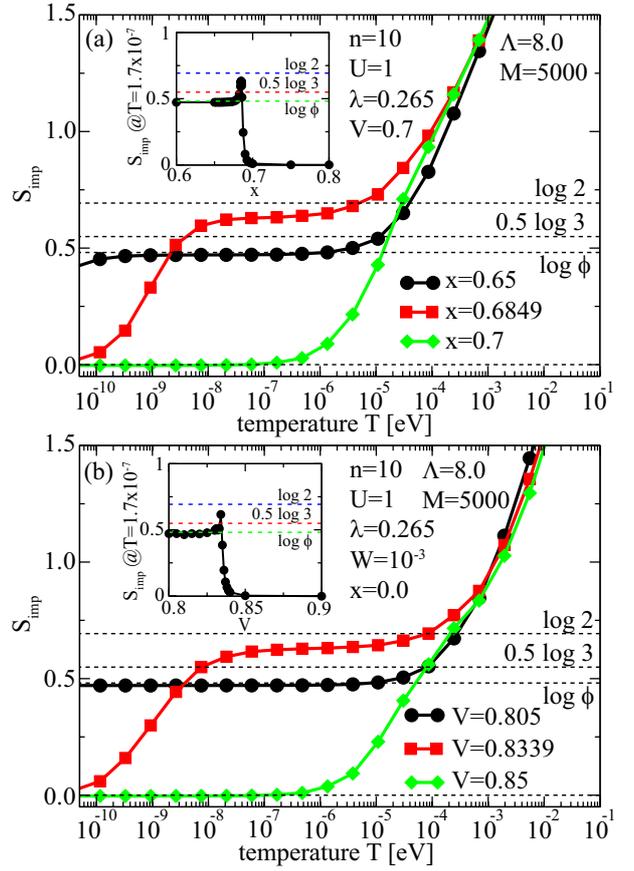}
\caption{(Color online)
(a) Entropies vs. temperature for $x=0.65$, $0.6849$,
and $0.7$ with $W=10^{-3}$ and $V=0.7$.
Inset denotes the residual entropies at $T=1.7 \times 10^{-7}$ vs. $x$
for $0.6 \le x \le 0.8$.
(b) Entropies vs. temperature for $V=0.805$, $0.8339$,
and $0.85$ with $W=10^{-3}$ and $x=0.0$.
Inset denotes the residual entropies at $T=1.7 \times 10^{-7}$ vs. $V$
for $0.8 \le V \le 0.9$.
}
\end{figure}

The QCP between the three-channel Kondo
and Fermi-liquid phases is expected to be characterized by
the residual entropy of the four-channel Kondo effect.
From eq.~(\ref{eq:Sana}), we obtain
$S_{\rm ana}=0.5 \log 3$ and $\log 2$
for $S=1/2$ and $1$, respectively, for $n_{\rm c}=4$.
Figure 3(a) shows the $f$-electron entropies
for $W=10^{-3}$ and $V=0.7$ between $x=0.65$ and $0.7$.
For $x=0.65$, we observe the three-channel Kondo phase,
while for $x=0.7$, the local singlet phase appears.
At $x=0.6849$, the entropy plateaus
with a value between $0.5 \log 3$ and $\log 2$,
and it exits the plateau at around $T \sim 10^{-10}$.
We deduce this as the signal of the QCP between
the three-channel Kondo and Fermi-liquid phases.
Note that this behavior has also been observed
in the region between $x=-0.7$ and $-0.6$.

In the inset of Fig.~3(a), we plot the value of the entropy
at $T=1.7 \times 10^{-7}$ as a function of $x$
between $x=0.6$ and $0.7$.
The entropy forms a sharp peak
at around $x \approx 0.685$, and the peak assumes
a value between $0.5 \log 3$ and $\log 2$.
Since this value is apparently larger than $0.5 \log 3$,
we deduce that the peak denotes the QCP characterized
by the residual entropy of the four-channel Kondo effect
with $S=1$.
This result is consistent with the considerations
of the local ground state.

Note that the entropy plateau does not reach $\log 2$,
which is the value obtained from eq.~(\ref{eq:Sana}).
The precision of
the NRG calculation provides a simple explanation for this.
If this is true, the plateau value is thought to converge to $\log 2$
for larger $M$ and smaller $\Lambda$.
Another explanation is the effect of the excited states on the entropy.
The QCP appears in the region where the local CEF state changes
between the $\Gamma_5$ triplet and $\Gamma_1$ singlet,
but for Ho$^{3+}$ ions,
there also exist $\Gamma_4$ triplet and $\Gamma_3$ doublet
excited states for a small excitation energy,
such as 26.9 K for $x=0.6849$ in Fig.~3(a).
Calculations of the NRG using larger $M$ and smaller $\Lambda$ values
must be performed in the future.
In addition, the possibility that the entropy plateau significantly
deviates from eq.~(\ref{eq:Sana}) owing to the effect of excited states
should be seriously considered.

To confirm the QCP behavior,
we also discuss the $V$ dependence of the entropy.
Figure 3(b) shows the $f$-electron entropies
for $W=10^{-3}$ and $x=0.0$ between $V=0.8$ and $0.85$.
For $V=0.805$, we observe the three-channel Kondo phase,
while for $V=0.85$, the local singlet phase appears.
At $V=0.8339$, we again encounter the QCP behavior.
Namely, an entropy plateau with a value
between $0.5 \log 3$ and $\log 2$ is observed.
In the inset of Fig.~3(b), we plot the value of the entropy
at $T=1.7 \times 10^{-7}$ as a function of $V$
between $V=0.8$ and $0.9$.
We observe that the entropy forms a sharp peak
at around $V=0.834$, and the peak assumes
a value between $0.5 \log 3$ and $\log 2$.
This peak is believed to denote the QCP characterized by
the residual entropy of the four-channel Kondo effect with $S=1$.

Here, we comment on the multipole state
to support the magnetic origin of $S=1$.
Since we included the three conduction bands,
multipoles up to rank 5 must be considered.
After the diagonalization of the multipole susceptibility matrix
obtained in the NRG calculations,
in the three-channel Kondo phase,
we found a $\Gamma_{4{\rm u}}$ multipole state
with the largest eigenvalue given by the combination
of dipole, octupole, and dotoriacontapole.
Note that ``u'' in the subscript denotes ungerade,
indicating the odd parity of time reversal symmetry.
Qualitatively, the $\Gamma_{4{\rm u}}$ multipole
state is consistent with the conclusion of the magnetic
three-channel Kondo effect characterized by $S=1$.
Details of the multipole susceptibility in the three-channel
Kondo phase will be discussed elsewhere in the future.

Finally, we briefly discuss 
the cubic Ho compound as a candidate for detecting
the three-channel Kondo behavior.
One way of investigating this compound
as a candidate to detect the three-channel Kondo behavior
is to synthesize HoT$_2$X$_{20}$
(T: transition metal element; X=Zn and Al).
Corresponding Pr and Nd compounds are considered to have
$\Gamma_3$ and $\Gamma_6$ ground states, respectively,
\cite{Onimaru1,Onimaru2,Higashinaka,Review,Nd1,Nd2}
suggesting that $W>0$ and $x \approx 0$
in the 1-2-20 crystal structure.\cite{Hotta1,Hotta2,Hotta2020}
Note also that the cage structure is useful for increasing the magnitude of
hybridization, because the rare-earth ion is surrounded by many ligand ions.
Thus, we expect the emergence of the three-channel Kondo
effect in HoT$_2$X$_{20}$.

In summary, we have investigated the seven-orbital Anderson model
hybridized with three conduction bands employing the NRG method.
We found a residual entropy of $\log \phi$,
which is characteristic of the three-channel Kondo effect in the wide region of
the local $\Gamma_5$ triplet ground state.
We also discussed the QCP between the three-channel Kondo
and Fermi-liquid phases.
We expect that the three-channel Kondo effect will be
observed in Ho 1-2-20 compounds.

The author thanks K. Hattori for discussions on the Kondo effect.
The computation in this work was partly done using the facilities of the
Supercomputer Center of Institute for Solid State Physics, University of Tokyo.


\begin{thebibliography}{99}

\bibitem{Nozieres}
Ph. Nozi\`eres and A. Blandin,
J. Phys. F {\bf 41}, 193 (1980).

\bibitem{Cox1}
D. L. Cox, Phys. Rev. Lett. {\bf 59}, 1240 (1987).

\bibitem{Cox2}
D. L. Cox and A. Zawadowski,
{\it Exotic Kondo Effects in Metals}
(Taylor \& Francis, London, 1999), p. 24.

\bibitem{Sakai}
A. Sakai and S. Nakatsuji,
J. Phys. Soc. Jpn. {\bf 80}, 063701 (2011).

\bibitem{Onimaru1}
T. Onimaru, K. T. Matsumoto, Y. F. Inoue, K. Umeo,
Y. Saiga, Y. Matsushita, R. Tamura, K. Nishimoto, I. Ishii, T. Suzuki,
and T. Takabatake, J. Phys. Soc. Jpn. {\bf 79}, 033704 (2010).

\bibitem{Onimaru2}
T. Onimaru, K. T. Matsumoto, Y. F. Inoue, K. Umeo, T. Sakakibara,
Y. Karaki, M. Kubota, and T. Takabatake, Phys. Rev. Lett. {\bf 106},
177001 (2011).

\bibitem{Higashinaka}
R. Higashinaka, A. Nakama, M. Ando, M. Watanabe, Y. Aoki, and H. Sato,
J. Phys. Soc. Jpn. {\bf 80}, SA048 (2011).

\bibitem{Review}
As a review, see also
T. Onimaru and H. Kusunose, J. Phys. Soc. Jpn. {\bf 85}, 082002 (2016)
and references therein.

\bibitem{Hotta1}
T. Hotta, J. Phys. Soc. Jpn. {\bf 86}, 083704 (2017). 

\bibitem{Fabrizio}
L. De Leo and M. Fabrizio, Phys. Rev. Lett. {\bf 94}, 236401 (2005).

\bibitem{spin-orbit}
S. H\"ufner, {\it Optical Spectra of Transparent Rare Earth Compounds},
(Academic Press, New York, 1978).

\bibitem{Hutchings}
M. T. Hutchings,
Solid State Phys. {\bf 16}, 227 (1964).

\bibitem{Slater}
J. C. Slater, {\it Quantum Theory of Atomic Structure}
(McGraw-Hill, New York, 1960).

\bibitem{Hotta-Harima}
T. Hotta and H. Harima,
J. Phys. Soc. Jpn. {\bf 75}, 124711 (2006).

\bibitem{LLW}
K. R. Lea, M. J. M. Leask, and W. P. Wolf,
J. Phys. Chem. Solids {\bf 23}, 1381 (1962).

\bibitem{NRG1}
K. G. Wilson, Rev. Mod. Phys. {\bf 47}, 773 (1975).

\bibitem{NRG2}
H. R. Krishna-murthy, J. W. Wilkins, and K. G. Wilson,
Phys. Rev. B {\bf 21}, 1003 (1980).

\bibitem{Hotta2005}
T. Hotta, J. Phys. Soc. Jpn. {\bf 74}, 1275 (2005).

\bibitem{Hotta2007}
T. Hotta, J. Phys. Soc. Jpn. {\bf 76}, 083705 (2007).

\bibitem{Affleck}
I. Affleck and A. W. W. Ludwig,
Nucl. Phys. B {\bf 360}, 641 (1991).

\bibitem{Koga1}
M. Koga and H. Shiba, J. Phys. Soc. Jpn. {\bf 64}, 4345 (1995).

\bibitem{Koga2}
M. Koga and H. Shiba, J. Phys. Soc. Jpn. {\bf 65}, 3007 (1996).

\bibitem{Kusunose1}
H. Kusunose and K. Miyake, J. Phys. Soc. Jpn. {\bf 66}, 1180 (1997).

\bibitem{Kusunose2}
H. Kusunose, J. Phys. Soc. Jpn. {\bf 67}, 61 (1998).

\bibitem{OSakai}
Y. Shimizu, O. Sakai, and S. Suzuki,
J. Phys. Soc. Jpn. {\bf 67}, 2395 (1998).

\bibitem{Koga3}
M. Koga, G. Zar\'and, and D. L. Cox,
Phys. Rev. Lett. {\bf 83}, 2421 (1999).

\bibitem{Koga4}
M. Koga, Phys. Rev. B {\bf 61}, 395 (2000).

\bibitem{Miyake1}
S. Yotsuhashi,  K. Miyake, and H. Kusunose,
J. Phys. Soc. Jpn. {\bf 71}, 389 (2002).

\bibitem{Fabrizio1}
M. Fabrizio, A. F. Ho, L. De Leo, and G. E. Santoro,
Phys. Rev. Lett. {\bf 91}, 246402 (2003).

\bibitem{Fabrizio2}
L. D. Leo and M. Fabrizio, Phys. Rev. B {\bf 69}, 245114 (2004).

\bibitem{Miyake2}
K. Hattori and K. Miyake, J. Phys. Soc. Jpn. {\bf 74}, 2193 (2005).

\bibitem{Koga5}
M. Koga and M. Matsumoto, Phys. Rev. B {\bf 77}, 094411 (2008).

\bibitem{Miyake3}
S. Nishiyama, H. Matsuura, and K. Miyake,
J. Phys. Soc. Jpn. {\bf 79}, 104711 (2010).

\bibitem{Miyake4}
S. Nishiyama and K. Miyake, J. Phys. Soc. Jpn. {\bf 80}, 124706 (2011).

\bibitem{Sela}
A. K. Mitchell and E. Sela, Phys. Rev. B {\bf 85}, 235127 (2012).

\bibitem{Shiina1}
R. Shiina, J. Phys. Soc. Jpn. {\bf 86}, 034705 (2017).

\bibitem{Shiina2}
R. Shiina, J. Phys. Soc. Jpn. {\bf 87}, 014702 (2018).

\bibitem{Hotta2}
T. Hotta, Physica B {\bf 536}, 203 (2018).

\bibitem{Koga6}
M. Koga and M. Matsumoto, J. Phys. Soc. Jpn. {\bf 88}, 034713 (2019).

\bibitem{Hotta3}
D. Matsui and T. Hotta, JPS Conf. Proc. {\bf 30}, 011125 (2020).

\bibitem{Hotta2020}
T. Hotta, J. Phys. Soc. Jpn. {\bf 89}, 114706 (2020).

\bibitem{Nd1}
Y. Yamane, R. J. Yamada, T. Onimaru, K. Uenishi, K. Wakiya, K. T. Matsumoto,
K. Umeo, and T. Takabatake, J. Phys. Soc. Jpn. {\bf 86}, 054708 (2017).

\bibitem{Nd2}
R. Yamamoto, T. Onimaru, R. J. Yamada, Y. Yamane, Y. Shimura, K. Umeo,
and T. Takabatake, J. Phys. Soc. Jpn. {\bf 88}, 044703 (2019).

\end{thebibliography}
\end{document}